# Programming Paradigms, Turing Completeness and Computational Thinking

Greg Michaelson[a]

a    School of Mathematical and Computer Sciences, Heriot-Watt University, Edinburgh, Scotland

**Abstract**    The notion of programming paradigms, with associated programming languages and methodologies, is a well established tenet of Computer Science pedagogy, enshrined in international curricula. However, this notion sits ill with Kuhn's classic conceptualisation of a scientific paradigm as a dominant world view, composed of a theory and complementary practice, which supersedes its predecessors through superior explanatory power.

  Furthermore, it is not at all clear how programming paradigms are to be characterised and differentiated. Indeed, on closer inspection, apparently disparate programming paradigms are very strongly connected. Rather, they should be viewed as different traditions of a unitary Computer Science paradigm composed of programming languages which are all Turing Complete, complemented by methodologies which may all be subsumed by Computational Thinking.

## The Art, Science, and Engineering of Programming

**Programming Paradigms, Turing Completeness and Computational Thinking**

## 1 Introduction

The number of different CPU designs in production has steadily decreased under competitive pressures, but a wide variety of deployed programming languages persists. Even though the use of some languages wanes until they effectively vanish, many stay in use for decades, alongside new languages that gain significant traction [32]. Similarly, the number of methodologies for designing programs from specifications seem resistant to competitive pressures: while the use of older methodologies declines, others persist, both integrated into and alongside new ones.

Given the ongoing variety of programming languages and methodologies, it is worthwhile seeking some organising principles, to enable understanding of how they may be distinguished and how they are related, to enable choice amongst them for particular problems, and to teach others how to best make such choices. Currently, the notion of different *programming paradigms*, with closely linked language classes and methodologies, is widely accepted as a basis for distinction.

For example, the object oriented (OO) paradigm might be seen as complementing specific OO languages like Java and C++ with an overarching OO methodology for identifying and manipulating objects, like UML. Similarly, the functional paradigm might be seen as complementing specific functional languages like Haskell and O'Caml with an overarching functional methodology comprising functional abstraction, higher order functions and recursive data structures.

This paper interrogates this idea of well characterised programming paradigms, and suggests that discriminating amongst them is fraught with problems. It is argued that apparently disparate languages are closely related, and that programming methodologies are not necessarily tied to particular classes of languages. Further, coexisting programming paradigms do not correspond to Kuhn's characterisation of a scientific paradigm as a dominant body of theory and practise. Instead, the language and methodology components of programming paradigms are closer to coexisting traditions within a unitary Computer Science paradigm, comprised of Turing Complete programming languages, and methodologies which may all be subsumed by Computational Thinking.

The paper is organised as follows:

- Section 2: defines and explains key programming language and methodology concepts, in particular strong Computational Thinking and Turing Completeness;
- Section 3: surveys historical approaches to classifying programming languages;
- Section 4: introduces Kuhn's notion of paradigm;
- Section 5: explores the emergence of the contemporary concept of a programming paradigm, and its place in international curricula;
- Section 6: analyses the canonical procedural, functional and object oriented programming paradigms, demonstrating that they are strongly linked in both language and methodology;
- Section 7: argues that Turing Completeness is the paradigmatic Computer Science theory, overarching programming languages;

- Section 8: argues that Computational Thinking, complementing TC languages, is the paradigmatic Computer Science practice, overarching programming methodologies;
- Section 9: draws conclusions and makes suggestion for future research.

## 2 Programming language and methodology concepts

In the following discussion, a *programming language* is a *symbol system capable of expressing computations*, and a *program* is *the expression of a computation in a programming language*. Here, "symbol" is interpreted liberally, to encompass any representation in any medium. In contrast, a "computation", however expressed, must have an equivalent sequence of operations for a Turing machine.

Then, a *programming methodology* is a way of *systematically designing a solution to a problem, from a specification, to a program which is executable on a digital computer*. Regardless of whether a specification is formal, in some mathematical notation, or informal, in some natural language, it must serve as a standard for assessing whether the program has solved the problem. While such assessment may in principle be by formal verification, in practice a combination of design inspection, empirical validation and socio-technical factors are the arbiters.

Finally, a *programming paradigm* is a *well characterised class of programming languages with a corresponding programming methodology*.

Note that, in this context, programming is concerned solely with the realisation of algorithms that meet specifications, and not with human facing aspects of computing such as interface and usability.

A central contention of this paper is that Computational Thinking (CT) is the paradigmatic methodology for Turing Complete (TC) computing. This will be explored in more depth below. To begin with, CT is defined and its use as a systematic discipline for programming is exemplified.

### 2.1 Computational Thinking

Following Wing's highly influential intervention [37], there has been sustained interest in CT, in particular as a basis for teaching programming. Michaelson [26] characterises a spectrum, from a weak CT, which sees pretty well all intellectual endeavour as CT, to a strong CT of systematic problem solving.

Strong CT is based on Kao's characterisation [21], which draws directly on Wing and is widely reproduced[1] The four elements of Kao's CT are:

- *Decomposition*: the ability to break down a problem into sub-problems;
- *Pattern recognition*: the ability to notice similarities, differences, properties, or trends in data;

---

[1] However, Kao is not commonly identified as the author, as the characterisation was developed by a group at Google that she chaired in 2010 (private email with David Harper and Elaine Kao, 2014).

- *Pattern generalization*: the ability to extract unnecessary details and generalize those that are necessary in order to define a concept or idea in general terms.
- *Algorithm design*: the ability to build a repeatable, step-by-step process to solve a particular problem. (p6)

Pattern generalisation may also be termed *abstraction*.

Note that algorithm design is programming language and paradigm neutral. Thus, algorithm design may be further distinguished from *coding*, which is the process of realising a design in a specific programming language.

## 2.2 Strong CT examples

CT may be used to systematically construct programs by interrogating concrete instances of problems. The following examples show the derivation of expressions with variables from sums, and of structured conditions from tabulated traces.

### 2.2.1 Finding variables

Variables may be introduced to generalise expressions, by comparison to identify the common pattern. For example, to find the area of a rectangle which is of height 4 cm and length 4 cm:

```
4 * 4 ==> 16
```

For example, to find the area of a rectangle which is 5 cm by 4 cm:

```
5 * 4 ==> 20
```

The two expressions may be compared to identify their shared structure, and hence where they differ:

```
4 * 4 vs 5 * 4 ==> ? * 4
```

That is, the common *pattern* has been located. A variable, say height, may now be introduced to *abstract* at the point of difference:

```
height * 4
```

### 2.2.2 Finding control structures

Control structures may be identified from traces of concrete computations by teasing out patterns and generalising them.

The basic structuring constructs common to all paradigms are:

- sequence e.g. command order, operator/procedure/function composition;
- choice e.g. conditional/case/guarded command/expression;
- repetition e.g. bounded/unbounded iteration/recursion.

For choice and repetition, actions are determined by differential properties of the data and state. Then patterns may be found in the connections between actions and conditions, and used to introduce structuring constructs. These may then be instantiated with abstracted expressions.

For example, suppose that a minibus tour operator wants to know how many buses are needed to satisfy all the reservations. The criteria are that reservations are first come, first served; a single reservation may request no more than 4 seats; each minibus has 12 seats; a minibus is deemed full once a reservation cannot be satisfied.

Suppose the queue for reservations requires the following numbers of seats:

3 4 5 2 4 6 3...

By hand, the concrete computation for this example, decomposed into data (requests), actions and conditions, proceeds as table 1. Counts for the number of minibuses so far and the number of seats taken in the current minibus have already been implicitly identified, so these may be made explicit.

◼ **Table 1** Concrete example computation

| requests | action | condition | minibuses | seats |
|---:|---|---|---|---|
| | | | 1 | 0 |
| 3 | reserve 3 seats | 4 or less seats requested and space in minibus | 1 | 3 |
| 4 | reserve 4 seats | 4 or less seats requested and space in minibus | 1 | 7 |
| 5 | reject booking | more than 4 seats requested | 1 | 7 |
| 2 | reserve 2 seats | 4 or less seats requested and space in minibus | 1 | 9 |
| 4 | start new minibus + reserve 4 seats | 4 or less seats requested and no space in minibus | 2 | 0 |
| | | | 2 | 4 |
| 6 | reject booking | more than 4 seats requested | 2 | 4 |
| 3 | reserve 3 seats | 4 or less seats requested and space in minibus | 2 | 7 |

Ignoring the sequence and counts, and grouping by condition, gives table 2.

◼ **Table 2** Concrete example computation grouped by condition

| requests | action | condition |
|---:|---|---|
| 3 | reserve 3 seats | 4 or less seats requested and space in minibus |
| 4 | reserve 4 seats | 4 or less seats requested and space in minibus |
| 2 | reserve 4 seats | 4 or less seats requested and space in minibus |
| 3 | reserve 3 seats | 4 or less seats requested and space in minibus |
| 4 | start new minibus + reserve 4 seats | 4 or less seats requested and no space in minibus |
| 5 | reject booking | more than 4 seats requested |
| 6 | reject booking | more than 4 seats requested |

These patterns of actions and conditions give the abstracted computation structure in listing 1.

The sums and conditions may be made explicit in the computation, as shown in table 3.

**Programming Paradigms, Turing Completeness and Computational Thinking**

■ **Listing 1** Abstracted computation structure

```
1  IF 4 or less seats requested THEN
2    IF space in minibus THEN
3      reserve seats
4    ELSE
5      start new minibus
6      reserve seats
7    ENDIF
8  ELSE
9    reject booking
10 ENDIF
```

■ **Table 3** Concrete example computation with explicit sums and counts

| requests | action | condition | minibuses | seats |
|---:|---|---|---:|---:|
| | | | 1 | 0 |
| 3 | add 3 to seats | 3 <= 4 and 0+3 <= 12 | 1 | 3 |
| 4 | add 4 to seats | 4 <= 4 and 3+4 <= 12 | 1 | 7 |
| 5 | reject booking | 5 > 4 | 1 | 7 |
| 2 | add 4 to seats | 4 <= 4 and 7+4 <= 12 | 1 | 9 |
| 4 | add 1 to minibus; set seats to 4 | 4 <= 4 and 9+4 > 12 | 2 | 4 |
| 6 | reject booking | 6 >= 4 | 2 | 4 |
| 3 | add 3 to seats | 3 <= 4 and 4+4 <= 12 | 2 | 7 |

■ **Listing 2** Initial algorithm

```
1  IF request <= 4 THEN
2    IF seats+request < 12 THEN
3      seats <- seats + request
4    ELSE
5      minibuses <- minibuses+1
6      seats <- 0
7      seats <- seats + request
8    ENDIF
9  ELSE
10   reject booking
11 ENDIF
```

Comparing concrete sums to identify patterns, and introducing variables to abstract differences, gives expressions which may be introduced into the computation structure to form an initial algorithm, as shown in listing 2.

Note that this algorithm may be further refactored, for example as listing 3.

The concrete computation in Figure 1 may also be analysed to identify an abstracted repetition of seeking and consuming more data, but this is not pursued here. For an example in the same style, see Michaelson[26].

▮ **Listing 3**  Refactored algorithm

```
1  IF request <= 4 THEN
2   IF seats+request > 12 THEN
3    minibuses <- minibuses+1
4    seats <- 0
5   ENDIF
6   seats <- seats+request
7  ELSE
8   reject booking
9  ENDIF
```

### 2.3  Turing Completeness and expressiveness

As Turing and Church hypothesised well before digital computers [8], all known models of computability may be shown to be TC, that is equivalent to Turing machines, through sound schema for translating instances of any one model into instances of any other. Similarly, all "full strength" programming languages may be shown to be equivalent, through schema for translating a program in any one language into any other. Furthermore, such languages are TC in that they may all express the same computations as a Turing machine, and hence any other model of computation. Ultimately, all executable code must necessarily be realised on a physical computer as machine code, either directly through compilation, or indirectly through an interpreter itself ultimately realised as machine code.

The primary requirement for a TC language is the capacity to describe unbounded computations over unbounded values. This encompasses a very wide spectrum of programming languages, if seemingly novel aspects are regarded as state changes, for example HTML manipulating WWW pages, and SQL manipulating databases.

Contrariwise, a non-TC language can only describe bounded computations, for example regular expressions, or manipulate bounded values, for example FORTRAN IV which, without files, has bounded numbers and arrays. Arbitrary instances of an arbitrary non-TC language will be translatable into an arbitrary TC language but not vice versa.

Orthogonal to Turing Completeness is Felleisen's formal notion of *expressiveness* [11]. Here one language is deemed less expressive than another if its semantics must be extended to make them equivalent. Thus, in Felleisen's terms, TC languages are equally expressive, and non-TC languages are less expressive than TC languages

### 2.4  Summary

This section has introduced the key programming language concepts used in the rest of the paper. Next, the notion of paradigm, and its application to programming languages and methodologies, will be explored.

## 3 Characterising languages

In the early days of Computing, it was usual to characterise languages by their "distance" from the bare CPU. Thus, *low level*, CPU specific machine codes and assembly languages were distinguished from platform oriented *autocodes*, in turn distinguished from platform independent *high level languages* [7]. This reflected a trade off between efficiency of resource use and ease of programming, given restricted language processor support. However, since the development of language processors targetting multiple platforms, with optimisation capabilities far exceeding that of human programmers, low level programming is now a niche specialism, and platform specific autocodes have been displaced by more general system programming languages.

It was also common for programming languages to be identified with distinct use domains. For example, IBM's Fortran was originally aimed at numerical applications, CODASYL's COBOL at data processing, MIT's LISP at symbolic computing for AI, and Bell Lab's C at system programming for UNIX. However, any TC language has the same expressive power as any other. Thus, while there may be pragmatic reasons for choosing a particular language for a particular domain, driven say by experience, language support, code base or customer requirement, in principle TC languages are interchangeable

The contemporary taxonomic approach uses formal properties of languages to identify distinct programming paradigms. These may include: memory model - imperative/mutable v declarative/immutable; type system - weak v strong, static v dynamic, ad hoc v parametric polymorphism; evaluation mechanism - strict v lazy, specified v unspecified order; abstraction mechanism - data structure, class, object, procedural, functional. Watt [35] provides a comprehensive overview of this approach, distinguishing imperative, object-oriented, concurrent, functional, logical, and scripting paradigms.

However, any one language may combine different or indeed overlapping combinations of these properties. For example, LISP may be viewed as both imperative and declarative. Watt [35] elides this by defining distinct paradigms, but then characterising some languages as *multi-paradigm*. Thus, C++ and Ada support both the object-oriented and imperative paradigms, enabling what he terms a hybrid programming style.

## 4 Kuhn's paradigms

Watt[34] was one of the first programming language textbooks to explicitly refer to the idea of a programming paradigm. The wider idea of a *scientific paradigm* derives from the work of the American philosopher Thomas Kuhn and his highly influential 1962 book "The Structure of Scientific Revolutions" [23].

For Kuhn, a *paradigm* is a way of conceptualising the world with a unitary body of theory and practice. Kuhn characterises *normal science* as the practice of the prevailing paradigm, and explores the processes underlying *paradigm shifts*, from an old to a new paradigm, in the face of explanatory failure. In turn, these new paradigms become the normal science. Kuhn explores a number of fundamental paradigm shifts, noting that they invariably take place against the intense resistance of practitioners of the prevailing normal science, who have social and personal interests in maintaining the status quo.

For example, in the 16th and 17th centuries, the Ptolemaic geocentric cosmology was displaced by the Copernican heliocentric cosmology. This was driven by new observations, enabled by new optical instruments, which could not be accounted for by the mathematics underpinning the geocentric cosmology. There was intense resistance to the new cosmology, in particular from the Catholic Church which saw heliocentrism as directly contradicting scripture. Indeed, the Church did not drop its proscription of Copernicus's and Galileo's books until 1835, despite the heliocentric model's universal adoption for astronomy and hence navigation.

The key property of a dominant scientific paradigm is that it can explain aspects of observable reality that prior paradigms cannot. For example, Newtonian physics cannot account for observable gravitational effects on light where relativistic physics can. While relativistic physics encompasses Newtonian physics, its is far more than simply an extension. Rather, Newtonian physics should be seen as a limiting case.

The Church-Turing hypothesis suggests that programming language components of different programming paradigms necessarily all have the same explanatory power. It is further argued below that Computational Thinking can capture the explanatory powers of the programming methodology components of programming paradigms. Thus, programming paradigms are not distinct in Kuhn's sense.

In Kuhn's conception, different paradigms do not and cannot coexist within normal science. One paradigm is always supreme until displaced by another. Nonetheless, Kuhn notes that the dominant paradigm is not uniform for all practitioners.

For example, while quantum mechanics is a unified body of knowledge, different people deploy it in different contexts. Thus, while an overall change to quantum mechanics will impact on all who use it, there may be local changes reflecting specialised application needs which do not have wider significance:

> ...though quantum mechanics (or Newtonian dynamics, or electromagnetic theory) is a paradigm for many scientific groups, it is not the same paradigm for them all. Therefore, it can simultaneously determine several traditions of normal science that overlap without being coextensive. (p52)

That is, different traditions that develop differentially from a common basis may coexist within normal science. It is argued below that this is a better characterisation of Computer Science than the prevailing notion of co-existing or hybrid paradigms.

## 5 The origins of programming paradigms

The contemporary idea of a programming paradigm may well have originated with Floyd's 1978 Turing Award Lecture [12] "The Paradigms of Programming".

Floyd's notion draws explicitly on Kuhn's. However, his focus is on programming as a problem solving methodology *supported by*, rather than *characterised by*, a language. Floyd identifies structured programming as the dominant Computer Science paradigm, which he presents as systematic stepwise refinement, from broad objective to complete code via structured constructs of sequence, iteration and condition, followed by information hiding, through procedural/functional and data abstraction. This is contrasted with branch-and-bound, divide-and-conquer and state-machine paradigms.

Floyd is insistent that language design should be driven by programming practice:

> *To the designer of programming languages, I say: unless you can support the paradigms I use when I program, or at least support my extending your language into one that does support my programming methods, I don't need your shiny new languages; like an old car or house, the old language has limitations I have learned to live with. To persuade me of the merit of your language, you must show me how to construct programs in it. I don't want to discourage the design of new languages; I want to encourage the language designer to become a serious student of the details of the design process.* (p459-50)

In 1986, Gibbs and Tucker [16], in their proposal for a new Computer Science curricula, explored the notion of *programming styles*, as part of the "Core Principles of Programming" topic. Here, they identify the procedural, functional programming, object programming, logic programming, modular programming and data flow, language styles. These are the familiar contemporary programming paradigms, reflecting the rapid changes in languages since Floyd, in particular the growth of OO and declarative programming. While these styles differ markedly from Floyd's problem analysis methodologies, as Floyd urges, the style drives the language.

Current usage had crystallised in the 1986 special issue of IEEE Software on Multi-paradigm Languages and Environments. In his Chief Editor introduction, Shriver [29], bemoans the absence of a common definition of software paradigm. For Shriver, in general:

> *a paradigm is a model or approach employed in solving a problem.* (p2)

He then characterises a software paradigm as being "induced by or inducing" a confluence of language class, programming environment and software engineering discipline. His example language classes include procedural, non-procedural, functional, visual, logic programming, and object-oriented. These intersect with, but do not correspond to, Gibbs' and Tucker's styles.

Finally, Shriver says that:

> *Software engineering refers to the set of rules and structured thinking that one applies within the environment to produce software systems or components.* (p2)

In his Guest Editor introduction to the same edition, Hailpern [18] expands on Shriver's distinction between *language-induced* paradigms and *paradigm-induced* languages. For language-induced paradigms, which historically came first, a programming paradigm is:

> *An abstract view of a class of programming that describes one means of solving programming problems.* (p8)

For the more recent paradigm-induced languages, a programming paradigm is:

■ **Table 4** Early programming paradigms
[ proc(edural); imp(erative); func(tional); par(allel); conc(urrent); alg(orithm)s]

| year | author | proc/imp | func | logic | OO | par/conc | algs | other |
|---|---|---|---|---|---|---|---|---|
| 1978 | Floyd [12] | | | | | | Y | |
| 1986 | Gibbs & Tucker [16] | Y | Y | Y | Y | | | modular data flow |
| 1986 | Shriver [29] | Y | Y | Y | Y | | | non-proc visual |
| 1986 | Hailpern [18] | Y | Y | | Y | Y | | access oriented data flow data structure oriented real time rules oriented |
| 1989 | Wegner [36] | | Y | Y | | Y | | distributed |

> *A way of approaching a programming problem. A way of restricting the solution set...A paradigm allows the programmer to use only restricted set of concepts (sic).* (p8)

Thus, for Hailpern:

> *Now that we can see the abstract paradigms behind the concrete languages, we can begin to design new languages based on these fundamental concepts.* (p8)

Hailpern's paradigms include access oriented, dataflow, data-structure oriented, functional, imperative, object oriented, parallel, real-time and rules-oriented.

In 1989, Wegner [36], in his Guest Editor introduction to a special issue of ACM Computing Surveys on programming paradigms, cites distributed, parallel, functional and logic programming. Here, the paradigms are based on commonalities in language constructs, that is, in Shriver's terminology, they are language-induced. Wegner identifies paradigms with classes of language.

This notion of paradigm encompassing a strong link between programming language and programming style still has considerable currency. It is this notion that is questioned here. Still, Wegner's comment remains pertinent:

> *Computer science has grown so fast, and fashions within computer science change so rapidly, that it is more difficult to identify its dominant paradigms or characterise its essence than in established scientific disciplines like physics or mathematics.* (p257)

Table 4 summarises programming paradigms in this period.

## 5.1 Paradigms in the curriculum

The Association for Computing Machinery (ACM) and the Institute of Electrical and Electronics Engineers (IEEE) are long-standing, international professional bodies based in the USA. In 1988, the ACM and IEEE Computer Society formed a Joint

Curriculum Task Force to develop a standard curricula for undergraduate Computer Science education. The Task Force was co-chaired by Allen Tucker whose work was discussed above, and who edited the 1991 Report [31].

This Report, and the 2011 [27] and 20013 [28] Revisions, have proved highly influential for both higher and secondary education curricula. However, each has deployed subtly different, and not always consistent, notions of programming and language paradigm.

The ACM/IEEE curricula has marked reach beyond North America. In the United Kingdom, post-school national benchmarks for curricula are specified by the independent Quality Assurance Agency for Higher Education (QAA). The Computer Science benchmark [14] explicitly follows the ACM/IEEE curricula. The British Computer Society accredits UK Computer Science courses, and its Guidelines are based on the QAA statement [6]. Such accreditation is accepted for wider UK Engineering Council, and hence European, chartered registration.

In English, Welsh and Northern Irish secondary schools, standards are defined by the UK Government's Department for Education. The mandated subject content for A Level Computer Science [13], the Year 13 exit qualification enabling University entrance, includes *"the need for and characteristics of a variety of programming paradigms"*(p2), without further qualification.

Examinations compliant with the standards are set by awarding bodies recognised by the government regulator. Schools may then choose amongst these. For example, the Assessments and Qualifications Alliance (AQA), an independent charity, is one of the five recognised bodies [1],

In Scottish secondary schools, curricula are defined by the Scottish Qualifications Authority (SQA). The Advanced Higher Computer Science [2], which is broadly equivalent to the A Level, species that knowledge and understanding of contemporary programming paradigms should be acquired.

Table 5 summarises the programming paradigms identified in these curricula.

## 5.2 Summary

There has been a transition in the notion of a programming paradigm, from problem solving methodologies, via their realisation in languages with appropriate constructs, to being characterised by languages with accompanying methodologies. Tables 4 and 5 summarise this transition, showing marked changes in the nature and scope of programming paradigms. Nonetheless, there appears to be agreement over the last 30 years that at least three core paradigms can be distinguished:

- *procedural programming*, also known as the *imperative* paradigm, based on modular decomposition and stepwise refinement, realised in "3rd generation" languages like Pascal and C;
- *object oriented programming*, based on class-oriented modelling, realised in OO languages like Smalltalk, C++, C#, Java and Python;
- *functional programming*, based on functional abstraction, realised in functional languages like LISP, Scheme, ML and Haskell.

▰ **Table 5** Paradigms in the curricullum
[ proc(edural); imp(erative); func(tional); par(allel); conc(urrent); alg(orithm)s]

| year | author | proc/imp | func | logic | OO | par/conc | algs | other |
|---|---|---|---|---|---|---|---|---|
| 1991 | ACM/IEEE [31] | Y | Y | Y | Y | Y | Y | |
| 2001 | ACM/IEEE [27] | Y | Y | | Y | | Y | distributed declarative scripting event driven |
| 2013 | ACM/IEEE [28] | Y | Y | | Y | Y | | scripting design: event driven, component level, data structure, aspect oriented & service oriented learning agents producer-consumer hardware |
| 2016 | AQA [1] | Y | (Y) | | Y | | | high v low level |
| 2016 | SQA [2] | Y | | | Y | Y | | |

## 6 Programming paradigms are not paradigmatic

On closer inspection, it is not clear how distinct these alleged paradigms actually are.

First of all, structured programming, based on algorithm elaboration using sequence, selection and repetition is an ambiguous case. It is a core procedural/imperative programming approach for expressing algorithms through stepwise refinement within and from modules.

Alternatively, , structured programming might constitute a paradigm in its own right; Floyd [12] thought it was the dominant paradigm. For example, it is about the only applicable methodology for initial programming in contemporary graphical languages, like Alice and Scratch, which rely heavily on jigsaw like pieces for sequence, choice and iteration as basic programming elements.

However, structured programming is a key methodology for OO as well as procedural programming, where it is used to elaborate algorithms within and from methods. Once classes have been identified, OO programming is indistinguishable from procedural programming. This suggests that structured programming is a subset of procedural programming, which, in turn, is a subset of OO programming.

Functional programming seems markedly different as there is no concept of assignment to modify mutable memory. However, structured programming is also a key methodology for functional programming, once functions have been identified: sequence is equivalent to function composition and repetition to recursion, all functional languages have explicit selection constructs, and modern varieties provide case based function definitions. Furthermore, modular decomposition and stepwise refinement are as applicable to functional as to procedural programming. This suggests that structured programming is a subset of procedural programming, which is a subset of both OO and functional programming - see figure 1.

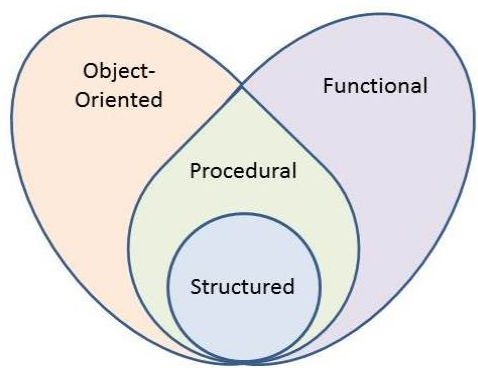

**Figure 1** Programming paradigms are not distinct

It is easier to realise a programming methodology in a language that directly supports corresponding constructs, and, in practice, a pre-specified language will strongly influence the programming approach. However, all of these methodologies are, in principle, programming language independent. For example, a design derived in a procedural methodology may be directly implemented in an OO language: much contemporary initial teaching is based on this. Similarly, an OO design may be implemented in a procedural language, with disciplined use of global sub-programs and data structures. Implementing strongly imperative designs in functional languages is more complicated, with copying substituting for in-place update. In contrast, implementing functional designs that do not take strong advantage of functional abstraction is straightforward in procedural or OO languages, which now invariably support recursion. Furthermore, functional abstraction may be realised through jump tables or classes.

## 7 Turing Complete computation as the paradigmatic Computer Science theory

Above, it was explained how, in the Church-Turing hypothesis, all TC languages are computationally equivalent, that is they all have the same expressive power. Arguably, TC computation is the only paradigm for the theory underpinning Computer Science. It was thought that analogue computing, based on differential equations, reified as differential analysers and then amplifiers, constituted a prior paradigm, but Babbage's early 19th century Analytic Engine [3] was TC, and analogue computing has been shown to be TC [17]. Similarly, quantum computing is widely heralded as disruptive but Deutsch [10] long established that it is TC.

Trans-finite hypercomputation would constitute a new paradigm by definition, as hypercomputers can allegedly solve decidability problems that Turing machines cannot, as, for example, Hogarth argues [20]. However, the possibility of hypercomputation is hotly contested [9].

Tedre [30] questions whether Computer Science is mature enough to even speak of paradigms. For Kuhn, one paradigm is dominant in any given epoch, until it is displaced by another paradigm with greater explanatory power. Thus, in Kuhnian terms, TC computation is the dominant theory for Computer Science, providing a substantial body of formalism underpinning the world changing practice of constructing and programming computers. This undermines Tedre's critique.

In terms of programming languages, it is not possible to identify clear disjunctions in language use; rather, new languages coexist with the old, which slowly fade away or are modified to more closely resemble the new. For example, object orientation has been added to procedural Fortran, LISP and C, and functional ML. Similarly, object oriented Java has been extended with functional abstraction.

There is a trend of convergence between TC languages, and choice seems increasingly driven by pragmatic considerations. Thus, from a programming language perspective, programming paradigms are akin to Kuhn's co-extant traditions, rooted within a common Computer Science theory of TC computation.

## 8 Computational Thinking as the paradigmatic Computer Science programming methodology

If TC computation is the paradigmatic theory of Computer Science, encompassing all TC programming languages, then there is strong case that CT is the complementary paradigmatic programming methodology. To establish this, it is necessary to show both that CT can account for other methodologies' disciplines, and that other methodologies cannot account for CT.

Above, a programming methodology was defined as a way of systematically designing a solution to a problem, from a specification, to a program which is executable on a digital computer. Note that this constrained conception of programming methodology is distinct from those concerned with aspects of programming in the broader socio-technical sense. For example, usability methodologies, such as User Centered Design or Third Wave HCI, address user focused specification and validation. For example, software construction methodologies, such as Agile or Live Programming, address implementation.

Michaelson [24] argues that many significant historical and contemporary programming methodologies can be encompassed in CT. Proponents of these methodologies have claimed that, like CT, they offer a systematic, disciplined way of deriving a program from a specification. The methodologies may be summarised in four classes as:

- programming language oriented: full strength language, pedagogic or full strength subset, functional, logic, object oriented, language independent/pseudo code;
- simple to complex: stepwise refinement, structured programming, iterative prototyping;

- components: modular programming, algorithms, data structures, types, classes, libraries;
- design: flowcharts, data flow, entity relationship, UML.

In CT terms, these methodologies are all variants of decomposition and algorithm design. However, only functional and logic programming, and type/data structure/class based methodologies, encompass pattern recognition and generalization.

While there are strong relationships within these methodology classes, it is not clear how any one methodology might capture the disciplines of all of the others, both within and across classes, including CT. UML comes closest to an overarching methodology, deriving from flow charts, data flow, ER and modular/classes, but does not address CT's pattern recognition and generalisation.

Formal methods, like calculational programming [5] and weakest pre-condition derivation [4], are forms of systematic stepwise refinement, with a strong focus on decomposition and algorithm construction, but have little to offer on pattern identification and generalisation. Similarly, approaches based on abstracted constructs like templates and design patterns [15] are sophisticated variants of structured programming, where the construct constrains the choice of arguments, as does a condition or iteration.

Overall, it is not possible to identify clear disjunctions in programming practice comparable to those from, say, Newtonian to relativistic mechanics. Rather, there is an evolution, where older programming practices build on and are absorbed into newer ones. For example, 70 years on from ENIAC, flowcharts are still used, and 1990s UML includes 1960's state machines. Thus, given that CT encompasses so many different programming methodologies, which it seems cannot encompass each other, in Kuhn's terms they may all be seen as representing complementary traditions within an overarching CT methodology, rather than as components of distinct paradigms.

## 9 Conclusion

It has been argued that, at present, the dominant Computer Science paradigm may be characterised theoretically as TC computation, overarching programming languages, and practically as Computational Thinking, overarching programming methodologies. This been couched in Kuhn's terms by reiterating the long established Church Turing hypothesis for the status of TC, and by suggesting that CT can account for contending methodologies which cannot account for CT. Thus, other methodologies constitute co-extant traditions that CT encompasses.

It is important to note that this argument does not depend on the actual uses of Turing machines or CT for programming. While TC is the standard term, any model of computation would suffice to characterise Computer Science's theoretical underpinning, including an arbitrary programming language with a well enunciated formal theory. Similarly, any methodology that can account for the others would suffice to characterise Computer Science practice. This paper suggests that CT is the strongest candidate to date.

Nonetheless, the assessment of properties of different programming languages remains indispensable for theoretical, practical and pedagogic practices. Formal language properties give strong and consistent taxonomic bases, such as those offered by Krishnamurthi [22] and Harper [19], who also question the notion of distinct programming paradigms tied strongly to distinct languages. Krishnamurthi focuses on a spectrum of language features through definitional interpreters, while Harper promotes understanding of programming languages in terms of types. Similarly, van Roy [33] uses a formal taxonomy, based on both operational properties and types, to identify thirty different programming "paradigms". While the conceptual status of these paradigms is moot, van Roy's systematic dissection of extant programming languages offers an extremely useful perspective.

The notion of distinct programming paradigms, as opposed to permeable traditions of programming languages and methodologies, now seems anachronistic in Computer Science curricula, and should be revisited. In particular, the inconsistent uses in the influential ACM/IEEE Curricula should be addressed.

In future research, it would be worth providing more detailed analyses of each of the programming methodologies considered above, through their actual deployment from specification to program, for a common exemplar, alongside the use of strong CT. This might falsify the assertion of CT's overarching status, or provide further evidence to support it.

**Acknowledgements** This paper grew from a seminar on Programming Paradigms, for Computer Science school teachers in the North West of England in December 2017. Subsequently, I summarised the seminar in a brief account for the magazine *(Hello World)* [25].

I would like to thank:
- Roger Davies, Paul Revell and participants at the CAS NW England HTP Day;
- Allen Tucker for stimulating discussion about the origins of programming paradigms;
- Andrew McGettrick and Mark Hogarth for help in locating references.

I would also like to thank the referees for their helpful critique.

## References

[1] Assessments and Qualifications Alliance. *AS AND A-LEVEL COMPUTER SCIENCE: AS (7516), A-level (7517)*. Version 1.4. AQA, December 2016. accessed 2020-01-12. URL: https://www.aqa.org.uk/subjects/computer-science-and-it/as-and-a-level/computer-science-7516-7517.

[2] Scottish Qualifications Authority. *Advanced Higher Computing Science Course Assessment Specification (C716 77)*. SQA, 2016. accessed 2020-01-13. URL: https://www.sqa.org.uk/files/nq/AHCourseSpec_ComputingScience.pdf.

[3] Charles Babbage. Passages from the Life of a Philosopher. In Phillip Morrison and Emily Morrison, editors, *Charles Babbage and his Calculating Engines*. Dover, 1961.

[4]  Roland Backhouse. *Program Construction: Calculating Implementations from Specifications*. Wiley, 2003.

[5]  Richard Bird and Oege de Moore. *The Algebra of Programming*. Prentice Hall, 1997.

[6]  British Computer Society. *Guidelines on course accreditation Information for universities and colleges*. BCS, January 2018. accessed 2020-01-12. URL: http://www.bcs.org/upload/pdf/2018-guidelines.pdf.

[7]  David G. Burnett-Hall, Leonard A. G. Dresel, and Paul A. Samet. *Computer Programming and Autocodes*. English Universities Press, 1964.

[8]  Paul Cockshott, Lewis M. Mackenzie, and Greg Michaelson. *Computation and its Limits*. OUP, 2012.

[9]  Paul Cockshott and Greg Michaelson. Are there new models of computation: a reply to Wegner and Eberbach. *Computer Journal*, 50(2):232–247, 2007. doi:10.1093/comjnl/bxl062.

[10]  David Deutsch. Quantum theory, the Church–Turing principle and the universal quantum computer. *Proceedings of the Royal Society A*, 400(1818), July 1985. doi:10.1098/rspa.1985.0070.

[11]  Matthias Felleisen. On the expressive power of programming languages. *Science of Computer Programming*, 17(1-3):35–75, December 1991. doi:10.1016/0167-6423(91)90036-W.

[12]  Robert W. Floyd. The Paradigms of Programming. *Computer Journal*, 22(8):455–460, 1979. doi:10.1145/359138.359140.

[13]  Department for Education. *GCE AS and A level subject content for computer science*. DFE-00359-2014. DFE, April 2014. accessed 2020-01-12. URL: https://assets.publishing.service.gov.uk/government/uploads/system/uploads/attachment_data/file/302105/A_level_computer_science_subject_content.pdf.

[14]  Quality Assurance Agency for Higher Education. *Subject Benchmark Statement: Computing*. QAA, February 2016. accessed 2020-01-12. URL: https://dera.ioe.ac.uk/25563/1/SBS-Computing-16.pdf.

[15]  Eric Gamma, Richard Helm, Ralph Johnson, and John Vlissides. *Design patterns: elements of reusable object-oriented software*. Addison-Wesley, 1994.

[16]  Norman E. Gibbs and Allen B. Tucker. A model curriculum for a liberal arts degree in Computer Science. *Communications of the ACM*, 29(3):202–210, 1986. doi:10.1145/5666.5667.

[17]  Daniel S. Graca and Jose S. Costa. Analog computers and recursive functions over the reals. *Journal of Complexity*, 19(5):644–664, October 2003. doi:10.1016/S0885-064X(03)00034-7.

[18]  Brent Hailpern. Guest Editor's Introduction Multiparadigm Languages and Environments. *IEEE Software*, 3(1):6–9, January 1986. doi:10.1109/MS.1986.232426.

[19]  Robert Harper. What, if anything, is a programming paradigm? *fifteeneightyfour blog, CUP*, May 2017. accessed 2020-01-12. URL: http://www.cambridgeblog.org/2017/05/what-if-anything-is-a-programming-paradigm/.

[20] Mark Hogarth. Non-Turing Computers are the New Non-Euclidean Geometries. *International Journal of Unconventional Computing*, 5(3-4):277–291, 2009. accessed 2020-01-30. URL: http://www.oldcitypublishing.com/journals/ijuc-home/ijuc-issue-contents/ijuc-volume-5-number-3-4-2009/ijuc-5-3-4-p-277-291/.

[21] Elaine Kao. Exploring Computational Thinking at Google. *CSTA Voice*, 7(2):6, May 2011. accessed 2020-01-14. URL: http://csteachers.org/documents/en-us/a0cf8f71-9771-41ec-b659-10cd6958fe3d/1/.

[22] Shriram Krishnamurthi. Teaching Programming Languages in a Post-Linnaean Age. *ACM SIGPLAN Notices*, 43(11):81–83, November 2008. doi:10.1145/1480828.1480846.

[23] Thomas S. Kuhn. *The Structure of Scientific Revolutions*. Chicago University Press, 1962.

[24] Greg Michaelson. Teaching programming with computational and informational thinking. *Journal of Pedagogic Development*, 5(1), March 2015. accessed 2020-01-14. URL: https://www.beds.ac.uk/jpd/volume-5-issue-01-march-2015/teaching-programming-with-computational-and-informational-thinking.

[25] Greg Michaelson. Are there programming paradigms? *(Hello World)*, 4:32–33, January 2018. accessed 2020-01-13. URL: https://helloworld.raspberrypi.org/issues/4.

[26] Greg Michaelson. Microworlds, Objects First, Computational Thinking and Programming. In *Computational Thinking in the STEM Disciplines: Foundations and Research Highlights*, pages 31–48. Springer, August 2018. doi:10.1007/978-3-319-93566-9_3.

[27] ACM/IEEE-CS Joint Task Force on Computing Curricula. *Computing Curricula 2001 Computer Science — Final Report*. ACM/IEEE-CS, December 2001. accessed 2020-01-13. URL: https://www.acm.org/binaries/content/assets/education/curricula-recommendations/cc2001.pdf.

[28] ACM/IEEE-CS Joint Task Force on Computing Curricula. *Computer Science Curricula 2013: Curriculum Guidelines for Undergraduate Degree Programs in Computer Science*. ACM/IEEE-CS, 2013. accessed 2020-01-12. URL: https://www.acm.org/binaries/content/assets/education/cs2013_web_final.pdf.

[29] Bruce D. Shriver. Software paradigms. *IEEE Software*, 3(1):2, January 1986. doi:10.1109/MS.1986.232422.

[30] Matti Tedre. *The Development of Computer Science: A Sociocultural perspective*. PhD thesis, University of Joensuu, 2006. accessed 2020-01-14. URL: https://core.ac.uk/download/pdf/15166913.pdf.

[31] Allen B. Tucker, Robert M. Aiken, Keith Barker, Kim B. Bruce, J. Thomas Cain, Susan E. Conry, Gerald L. Engel l, Richard G. Epstein, Doris K. Lidtke, Michael C. Mulder, Jean B. Rogers, Eugene H. Spafford, and A. Joe Turner. *Computing curricula 1991: Report of the ACM/IEEE-CS Joint Curriculum Task Force*. ACM/IEEE-CS, 1991. doi:10.1145/126633.

[32] Liam Tung. Programming language popularity: C++ bounces back at Python's expense. *ZDNet*, April 8 2019. accessed 2020-01-12. URL:

      https://www.zdnet.com/article/programming-language-popularity-c-bounces-back-at-pythons-expense/.

[33] Peter van Roy. Programming Paradigms for Dummies: What Every Programmer Should Know. In Gerard Assayag and Andrew Gerzso, editors, *New Computational Paradigms for Computer Music*, pages 9–47. IRCAM/Delatour, France, 2009.

[34] David A. Watt. *Programming Language Concepts and Paradigms*. Prentice Hall, 1990.

[35] David A. Watt. *Programming Language Design Concepts*. Wiley, 2004.

[36] Peter Wegner. Guest Editor's Introduction to Special Issue of Computing Surveys on Programming Language Paradigms. *ACM Computing Surveys*, 21(3):253–258, September 1989.

[37] Jeanette M. Wing. Computational Thinking. *Communications of the ACM Viewpoint*, pages 33–35, March 2006. doi:10.1145/1118178.1118215.

**About the author**

**Greg Michaelson** is Emeritus Professor of Computer Science at Heriot-Watt University. His research interests are in the design, implementation and analysis of programming languages, in particular functional languages for multi-processor deployment. You can find his work at: https://www.macs.hw.ac.uk/~greg and may contact him at: G.Michaelson@hw.ac.uk. 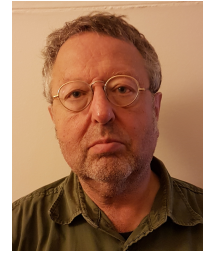